\documentclass{revtex4}
\begin{document}
\begin{center}
{ \huge \bf
Bi-layer-dimerized chiral liquid crystals.}
\end{center}
\vskip1cm
\begin{center}
{\huge   M. Hudak\\{\Large Stierova 23, SK-0401 Kosice} } \\
\vskip0.5cm
{\huge O. Hudak\\{\Large  Department of Aviation Technical Studies , Faculty of Aerodynamics, Technical University Kosice, Rampova 7, SK-04001 Kosice} } 

\end{center}

\maketitle

\section*{Abstract}
There is a large variety of bi-layered structures with smectic A
type ordering. Results of this
paper contribute to a theoryof bi-layered phases.
 The difference between the \( SmC^{*}_{A} \) and
\( SmC_{\alpha}^{*} \) phases is described in this latter paper and
it is explained by existence and non-existence of the dipole
moment pairs.
\section{Introduction}

In \cite{111}  authors discuss several types of bilayer smectic liquid crystals with ferroelectric and antiferroelectric properties in binary mixture of dimeric compounds.
The mesomorphic behavior and phase structure were examined in \cite{111} the mixture of two kinds of dimeric compounds, α,ω-bis(4-alkoxyanilinebenzylidene-4‘-carbonyloxy)pentane (mOAM5AMOm), by optical microscopy, X-ray diffraction, polarization switching, and second-harmonic generation measurements. One compound is 4OAM5AMO4 with a short terminal alkyl chain that forms a single-layer smectic phase (SmCAs) with a random mixing of spacer and tail groups. Another compound is 16OAM5AMO16 with a long terminal alkyl chain that forms a chiral, anticlinic, and antiferroelectric bilayer phase (SmCAb) with the bent molecules tilted to the bilayer. By mixing these two compounds, the SmCAs phase of 4OAM5AMO4 is easily destabilized, leading to the wide content region of the bilayer phases. In the bilayer regime, three other smectic phases are newly induced. Two of them are antiferroelectric and ferroelectric phases in which the molecules lie perpendicularly with respect to the layer. The other shows no polar response to an external electric field and behaves like a smectic A. The new appearance of these bilayer phases is discussed as a mixing effect of long and short tail groups. Thus it is useful to discuss bi-layer-dimerized chiral liquid crystals theoretically.

Let us note that liquid crystal television display panel has application f.e. in tactical aircraft. Its key advantages are: (1) high contrast in small and large areas, (2) shade capability under all levels of illumination including direct-sunlight, (3) uniform high resolution over the entire display area, (4) interface similar to CRT TV display and (5) low power, weight, volume. Cockpit installations have been designed for the display which permit viewing under day and night conditions, see in \cite{112}.

Isotropic liquids represent the most symmetric
phase: there is no discrete translational symmetry, the local point group
symmetry is present only.
Usually the point group in liquids is
identified with the point symmetry of a group of molecules forming
the unit, from
which the liquid or liquid crystal is formed.

In the isotropic symmetry the unit
has
either full orthogonal symmetry group \( K_{h} \), either a group
\( K \) of
rotations around a point where inversion is absent.

A phase
transition due to symmetry lowering  of the isotropic liquid may
occur, usually there are transitions to the nematic or the
Sm A (centrosymmetric, nonchiral) phases \( ( K_{h} \rightarrow D_{\infty
h} ), \) Sm A (without the inversion center, chiral) \( ( K \rightarrow D_{\infty}
), \)
and to cholesteric phases. Symmetry groups \( D_{\infty} \) and \(
D_{\infty h} \) correspond to all \cite{P} known uniaxial liquid
crystals with those phases. These liquid crystals are characterized by
the fact that both directions along the axis of the full axial
symmetry are equivalent.
The director {\bf n} orientation
is equivalent to the opposite director orientation -{\bf n}.

The order parameter for these liquid crystals may be found using the
well-known
tensor \( Q_{ij}: \)
\[ Q_{ij}({\bf r}) = <l_{i} l_{j} > - \frac{1}{3} <l_{k}l_{k}>
\delta_{ij}, \]
which is formed as the local average of the quadratically mixed
long molecular unit axis {\bf l} projections.
It is possible to rewrite the above order parameter expression
into the form
\[ Q_{ij}({\bf r}) = Q({\bf r}) (n_{i}({\bf r}) n_{j}({\bf r})
- \frac{1}{3}  \delta_{ij}). \]
Here the tensor amplitude Q({\bf r}) tells us in which extent
the molecules, forming the unit, are aligned  in the direction
given by the director {\bf n} (and equivalently {\bf -n}).

Long molecules of lower molecular weight may have permanent
electric dipole moments which are generally oriented
to the long
molecular axes. Ferroelectricity due to their alignment  is not
observed due to several main reasons \cite{FTIIT} :
\begin{enumerate}
\item dipole-dipole interaction is weak and the thermal energy \(
k_{B} T \) destroys such an order,
\item when the molecular dipole moments are large, the fluidity
leads to formation of dimers, in which an antiparallel
orientation of the molecular dipoles exist, thus the effective
dipole moment is canceled,
\item the flexoelectric effect may lead to deformation which as a
consequence is relaxed by the formation of defects.
\end{enumerate}

A model which is widely accepted for the antiferroelectric
smectic phases suppose that
the molecules in the neighbouring layers tilt in the
opposite directions, and so also the polarization point in the
opposite directions.
Two
vectors \( \theta_{1} \) and \( \theta_{2}, \) representing
orientations of molecules in the alternating layers, may be
defined
\[ \theta_{1,2} = (-n_{1,2 y}n_{1,2 z},n_{1,2 x}n_{1,2 z}). \]
where \( {\bf n}_{1,2} \) are the directors in the odd-numbered
and even-numbered layers, respectively. The axis z is perpendicular to the
layer.

Such a model assumption corresponds with expectation that there are
weaker {\bf interlayer} correlations between molecules while
{\bf inlayer} interactions are stronger. These latter
include also interactions leading to in-layer dimerization.
Dipole-dipole forces may however strongly influence also interactions
between molecules localized in different neighbouring layers.
Then the interlayer dimerization between
neighbouring layers may occur.
Competition between the in-layer
and inter-layer dimerization may be influenced by such factors as
the in-layer space restrictions for molecular grouping and the
value of the molecular dipole moment.
In extreme the interlayer dimerization processes supported
by the in-layer geometry restrictions may be leading forces
driving formation of the layered liquid crystal structure when
cooling the isotropic liquids.

It is this situation, in
which pairing dimerization occurs predominantly between molecules
on both sides of a layer interface, which we concern  with in this
paper. In traditional approach localization of these dimers is considered
to be in the middle of every layer.

Due to interlayer pairing of molecules we expect that
the invariance
of the crystal with respect to the change of the direction {\bf n}
to {\bf -n} in every layer is broken.
Molecular groups
(dimers), which are
invariant with respect to the inversion are situated between two
molecular layers . 

One may expect that this new kind of liquid crystal ordering in
which inter-layer dimerization prevails may, under appropriate conditions,
occur when decreasing temperature of the isotropic liquid. In this
case
in the isotropic liquid tend to correlate more and more
their movements. This process may be characterized by
two correlation
lengths: that of the above mentioned
dimerization process and that of the interactions leading to a
layered structure. Usually
the characteristic dimerization length has been expected to be
smaller or comparable
with the interlayer forces the characteristic length of which is
approximately equal to the interlayer spacing. If,
however, the latter is smaller than the former then the
inter-layer dimerization
becomes realistic; i.e. dimerization groups centers are
localized at the boundaries of neighbouring layers.

Our aim in the following part of the paper
is to construct appropriate theory describing ordering
and transitions just described, and to discuss experimental
consequences  and
observations within the frame of this theory.

\section{Order parameter}
As a consequence of our expectation that molecular "dimer" groups
are preferentially   formed between two neighbouring smectic
layers of molecules, two
vectors \( \theta^{'}_{1} \) and \( \theta^{'}_{2}, \) representing
orientations of molecules in the alternating
neighbouring layers, may be
defined
\begin{equation}
\label{1}
\theta^{'}_{1} = (-n_{1, y}n_{2, z},n_{1,x}n_{2, z}),
\theta^{'}_{2} = (-n_{2 y}n_{1,z},n_{2,x}n_{1,z}),
\end{equation}
where \( {\bf n}_{1,2} \) are the directors in the odd-numbered
and even-numbered layers, respectively. The axis z is perpendicular to the
layer.

Symmetric ({\it s}) and antisymmetric ({\it a})
combinations of both vectors may be
formed:
\begin{equation}
\label{2}
s \equiv   \theta^{'}_{1}+\theta^{'}_{2},
a \equiv   \theta^{'}_{1}-\theta^{'}_{2}.
\end{equation}

While {\it s} does not change its sign under a basic translational
symmetry: transformation shifting layers by one layer thickness,
it is not the
case of the antisymmetric combination {\it a}. The quantity {\it
s} has
symmetry properties of the  in-bilayer
ferroelectric polarization. The latter quantity
represents the in-bilayer antiferroelectric polarization.

Compare  axial vectors \( \theta^{'} \) from (\ref{1}) with those
which are usually introduced \cite{LBT} for traditional
antiferroelectric smectic phase description:
\begin{equation}
\label{3}
\theta_{1} = (-n_{1, y}n_{1, z},n_{1,x}n_{1, z}),
\theta_{2} = (-n_{2 y}n_{2,z},n_{2,x}n_{2,z}).
\end{equation}
Both types of the mentioned axial vectors become the same
quantity whenever the ordering of layers one and two is the same.
However, they are different if
the geometrical and dipolar forces
tend to dimerize groups of molecules in adjacent layers.

Another comparison, results of which are displayed in Table 1,
\begin{table}
\centering
\label{T}
\caption{Different behaviour of the smectic layered
phases under inverse transformations.}
\vspace{0.1in}
\begin{tabular}{||l|c|c|c|c|c||} \hline
layer & 1 & 1 & 1 & i & i \\ \cline{2-6}
      & 2 & 1 & i & 1 & i \\ \hline
layered ordering type & \cite{LBT} & 1 & 1 & 1 & 1 \\ \cline{2-6}
                      & this paper & 1 & -1 & -1 & 1 \\ \hline
\end{tabular}
\end{table}

shows different behaviour of the bilayer vectors {\bf \it s} and
{\bf \it a}
when the inverse transformations i ( \( {\bf n}_{1,2} \rightarrow
{\bf -n}_{1,2}\) ) are performed in all four combinations
for two adjacent layers:
layers 1 and 2 remained unchanged,
the layer 1 inverted only (the remaining layer 2 is unchanged),
the layer 2 inverted only (the remaining layer 1 is unchanged),
and the
layers 1 and 2 simultaneously inverted.

In the Table 1 we denote by 1 and by i
the identity and the in-layer inverse transformations. The third
and fourth
rows contain results of this transformations on vectors {\bf \it
s} and {\bf \it a} obtained
when the ferroelectric and antiferroelectric phases are defined
by the way used in \cite{LBT} and in this paper. Symbol 1 denotes
resulting
identity transformation in the axial vectors s, a space:
\( (s \leftrightarrow s) \) and
\( (a \leftrightarrow a) \) ; and symbol -1 represents result of inverse
transformation in the axial vectors s,a space:
\( (s \leftrightarrow -s) \) and
\( (a \leftrightarrow -a). \)

The most pronounced difference between our new order parameters
and the traditional ones becomes apparent whenever phases with
modulation of the ground state in the direction perpendicular to
the smectic layers occur.
If this modulation is weak and with
long wavelength with respect to the interlayer distance a, then
one can relate director vectors in neighbouring layers:
\begin{equation}
\label{4}
n_{2,\alpha} = n_{1, \alpha} + a \frac{\delta n_{1,
\alpha}}{\delta z}, \nonumber \\
n_{1,\alpha} = n_{2, \alpha} - a \frac{\delta n_{2,
\alpha}}{\delta z}.
\end{equation}
Our new order parameters \( \theta^{'}_{1} \) and \( \theta^{'}_{2}\)
are related to the well-known order parameters \( \theta_{1} \) and \( \theta_{2}: \)
\begin{equation}
\label{5}
\theta^{'}_{1} = \theta_{1} + a(-n_{1,y} \frac{\delta n_{1,\alpha}}{\delta z}, n_{1,x} \frac{\delta n_{1,\alpha}}{\delta z}), \nonumber \\
\theta^{'}_{2} = \theta_{2} - a(-n_{2,y} \frac{\delta n_{2,\alpha}}{\delta z}, n_{2,x} \frac{\delta n_{2,\alpha}}{\delta z}).
\end{equation}
Both sets of order parameters \( \theta \) and \( \theta^{'} \) become identical, whenever there is no gradient change of the
director orientation from layer to layer. If this modulation is not weak then one can relate director vectors in neighbouring layers
using finite differences of directors:
\begin{equation}
\label{6}
n_{2,\alpha} = n_{1, \alpha} + \delta ,
\end{equation}
and obtains, that our new order parameters \( \theta^{'}_{1} \) and \( \theta^{'}_{2} \)
are related to the order parameters \( \theta_{1} \) and \( \theta_{2}: \)
\begin{equation}
\label{6'}
\theta^{'}_{1} = \theta_{1} + (-n_{1,y} \delta , n_{1,x} \delta),
\nonumber \\
\theta^{'}_{2} = \theta_{2} - (-n_{2,y} \delta , n_{2,x} \delta
).
\end{equation}
Both sets of order parameters \( \theta \) and \( \theta^{'} \) become identical, whenever the finite difference \( \delta \) vanishes.

Thus we conclude, that our order parameters are more general than those traditionally
used, the latter mentioned are special realizations  of the former.

\section{Free energy}
Up to the fourth order the free energy F expansion in small
amplitudes of the order parameters {\it s} and {\it a} has the
form
\begin{equation}
\label{7}
F= \frac{\alpha}{2}I_{1}+\frac{a}{2}I_{2}+cI_{1}I_{2}+dI_{3},
\end{equation}
due to the fact that there are only the following
second order  invariants
\[ I_{1}= s^{2},  I_{2}= a^{2}, \]
and the fourth order invariants
\[ I_{3}= (s.a)^{2}, I_{1}I_{2}, I^{2}_{1}, I^{2}_{2}. \]
There are five ordered low-symmetry phases: the
ferroelectric phase
\begin{equation}
\label{8}
s \not= 0,  a = 0;
\end{equation}

the antiferrolectric phase
\begin{equation}
\label{9}
a \not= 0,
s = 0;
\end{equation}
the ferrielectric phase A \( (a \parallel s) \)
\begin{equation}
\label{10}
a \not= 0,
s \not= 0;
\end{equation}
the ferrielectric phase B \( (a \perp s) \)
\begin{equation}
\label{11}
a \not= 0,
s \not= 0;
\end{equation}
the ferrielectric phase C \( (0< a.s < \vert a \vert . \vert s
\vert) \)
\begin{equation}
\label{12}
a \not= 0;
s \not= 0.
\end{equation}

Until  now we did not consider any difference between ordering in
which both neighbouring layers are ordered in the same way
(uni-layer-ordering) and
ordering in which neighbouring layers are ordered in a different
way (bi-layer-ordering). Our free energy (\ref{7}) does not
contain a term which breaks the symmetry between bi-layer-ordering and
uni-layer-ordering. This term will be discussed latter on.

The ordering of molecules which corresponds to these
five phases can be conveniently described introducing usual spherical coordinates \( \phi \)
and \( \Theta \) in each
individual layer
\begin{equation}
\label{4'}
{\bf n}_{i} = (cos(\phi_{i}) sin(\Theta_{i}),sin(\phi_{i})
sin(\Theta_{i}), cos(\Theta_{i})).
\end{equation}

Note, that while the azimuthal angle \( \phi \) takes values
from 0 to \( 2 \pi \) (excluding the last value), the tilt angle
takes values
from 0 to \( \pi \) (excluding the last value) due to absence of the
inversion in-layer symmetry.

Let us discuss all possible molecular ordering corresponding to
individual phases. For the ferroelectric phase (\ref{8})
we obtain from the condition a=0 that the rotation and tilt angles
for the layer 1 and the layer 2 are related in this phase
via the following equations:
\[ sin(\phi_{1}) sin(\Theta_{1}) cos(\Theta_{2}) =
 sin(\phi_{2}) sin(\Theta_{2}) cos(\Theta_{1}) \]
and
\[ cos(\phi_{1}) sin(\Theta_{1}) cos(\Theta_{2}) =
 cos(\phi_{2}) sin(\Theta_{2}) cos(\Theta_{1}). \]
In correspondence with the mentioned double
degeneracy there are two possibilities how to satisfy these equation
conditions. The first one is that in which there is
equal rotation angle in both layers \( \phi_{1} = \phi_{2},
\) and equal tilt angle in both layers \( \Theta_{1} = \Theta_{2}
\). This state corresponds to
the usually considered
description of the smectic ferroelectric phase.
In the other one the rotation angles in neighbouring  adjacent
layers are differing by 180 degrees \( \phi_{1} = \phi_{2}
\pm \pi, \) and  tilt angles in both layers are complementary to
180 degrees differing \( \Theta_{1} = \pi - \Theta_{2} \).
This state does not corresponds to any traditionally accepted
ordering of the smectic ferroelectric phase. It is a new
state, let us name this state as bi-layered-ferroelectric state.
Both types of the ferroelectric states are schematically visualized in Fig.1.

In the antiferrolectric phase
the condition s=0, ref{9}, holds and
we obtain that the rotation and tilt angles
for the layer 1 and the layer 2 are related in this phase
via the following equations:
\[ sin(\phi_{1}) sin(\Theta_{1}) cos(\Theta_{2}) =
- sin(\phi_{2}) sin(\Theta_{2}) cos(\Theta_{1}) \]
and
\[ cos(\phi_{1}) sin(\Theta_{1}) cos(\Theta_{2}) =
- cos(\phi_{2}) sin(\Theta_{2}) cos(\Theta_{1}). \]
There are again two possibilities how to satisfy these equation
conditions. The first one is that in which
rotation angles in neighbouring  adjacent
layers are differing by 180 degrees \( \phi_{1} = \phi_{2}
\pm \pi, \) and  tilt angles in both layers are complementary to
180 degrees also
differing \( \Theta_{1} = \Theta_{2} \).
 This state corresponds to the usually accepted
description of the smectic antiferroelectric phase.

The other one is characterized by
equal rotation angle in both layers \( \phi_{1} = \phi_{2},
\) and equal tilt angle in both layers \( \Theta_{1} = \pi - \Theta_{2},
\).
This state does not corresponds to any traditionally considered
smectic antiferroelectric phase. It is a new
state, let us name this state as bi-layered-antiferroelectric state.
Both possibilities in which antiferroelectric ordering realizes
are schematically visualized in Fig.2.

The ferrielectric phases, (\ref{10}) - (\ref{12}), A where \( (a \parallel
s), \) B where \( (a \perp s), \)
C where \( (0< a.s < \vert a \vert . \vert s \)
may be similarly discussed as ferroelectric and antiferroelectric
phases above.

It is possible to discuss phenomenological theory of ordering
transitions between just described phases, where
also inhomogeneous terms contributing to the free energy play
important role.
Such a description
will be made in another paper together with discussion of
behaviour of relevant quantities.

Here we concentrate our attention on the bi-layered
antiferroelectric phases. Our aim is to discuss mechanisms which,
on the semiphenomenological level, lead to bilayered structures.
In the next section we develop a
semiphenomenological theory of the McMillan type which enables
to obtain better insight how new kind of antiferroelectric
ordering discussed above may evolve from the isotropic phase.
The case of ferroelectric and
ferrielectric phases is not considered in this paper in order
to simplify illustration of our model assumptions and
consequences .

\section{Generalized McMillan theory of strongly dipolar
liquids of the Sm A type}

In this section we discuss a generalized form of the McMillan
theory of strongly dipolar liquids. In its original form it is
describing liquids in which
there are
weak inter-layer correlations between molecules while
strong  in-layer interactions are responsible for the
mechanism of the molecular ordering with
prevailing in-layer dimerization. The theory presented below is our
generalization of the McMillan version \cite{SCh} of the mentioned theory
for traditional type of Sm A ordering
described in to those materials in which dimerisation
effects between neighbouring layers are strong. McMillan
generalized the Maier-Saupe theory introducing a new order
parameter, which characterizes one dimensional translational
periodicity of the layered smectic structure. In the next section
we generalize  this McMillan theory by introducing another new
order parameter, which characterizes bi-layered ordering due to
strong dipolar interactions resulting in interlayer dimerisation
of molecules.

\subsection{The effective potential}
Anisotropic part of the pair interaction potential is assumed to
have the form
\begin{equation}
\label{P}
V_{12}=- \frac{V_{0}}{2r^{3}_{0}\sqrt{\pi}}
\exp{(-\frac{r^{2}_{12}}{r^{2}_{0}})} (3 \cos^{2}(\theta_{12})-1)
\end{equation}
where short range character of the interactions is described by
the exponential part of the potential and
where \( {\bf r}_{12}={\bf r}_{1}-{\bf r}_{2} \)
is the distance between centers of two interacting molecules 1
and 2, \( \theta_{12}=\theta_{1}-\theta_{2} \)
is the angle between long axis of these molecules,
and where \( V_{0} \) is the interaction potential energy
constant, \( r_{0} \) is a characteristic molecule
length,  \( \theta_{1,2} \)  and \(
{\bf r}_{1,2} \) are tilt angles and positional vectors of the molecules 1 and 2.
The potential (\ref{P}) leads, adopting steps similar as in
\cite{SCh}, to a
semiphenomenological form of
the effective potential describing interaction of a given molecule
with the effective medium in layers of the thickness a:
\begin{equation}
\label{EP}
V=-V_{0}(\Sigma+ \sigma \alpha \cos(\frac{2\pi z}{a})) \frac{1}{2}
(3 \cos^{2}(\Theta)-1),
\end{equation}
where
\( V_{0} \) is the phenomenological energy
constant, z is a coordinate in the ordering direction, \( \alpha \) is
a phenomenological constant describing relative strength of energy scales
for nematic and smectic ordering, \cite{SCh}, \( \alpha = 2 \exp(-(\frac{\pi r_{0}}{a})^{2}),
\) \( \theta \) is a tilt angle of a given molecule with respect to
the effective medium, \( \Sigma
\) is the
nematic order parameter, \( \sigma \) is the smectic order
parameter.

For \( \Sigma \) nonzero
this potential prefers nonlayered nematic-type
dipole ordering.
For \( \sigma \) nonzero layered ordering of the is smectic type
is prefered.
In the Fig.3 the most
advantageous orientations of long molecules (dimerization) in the
traditional, \cite{P},  case (I.) and in the case considered here (II.)
are shown.

To model bi-layered ordering we
assume that there exists an additional anisotropy potential due to dipolar forces
in neighbouring layers and due to geometrical restrictions in
layered ordering.
When constructing that potential we have in mind that
strongly polar molecules are
characterized  by the short range interactions of the Van
der Waals type and by dipolar
interactions. While the first type of interactions is already
included in the form of the effective
potential (\ref{EP}), the latter is not. The effective medium, which
interacts with the given molecule is formed by the
nonpolar background and a polar group of molecules in the
neighbouring layer. This latter interaction may be described by
the same way as in nematics when studying  the
antiferroelectric short range order \cite{SCh}. Thus it is
expected to be proportional to \( \cos(\Theta). \)  In other
words we expect that there is a tendency to orient the molecule
head  with respect to the neighbouring layer.
To describe the layer forming tendency
we expect, that a term \( \cos(\frac{2 \pi z }{2a}) \) will be the
most important term. Note that in this term there is the length of
two layer separation 2a  due to the fact that we
expect prevailing orientation in a smectic layer with the
molecular heads up (down) and
in the next layer down (up). Namely this bi-layered
structure has periodicity 2a. Thus summarizing  we expect that a
new effective potential term has the form
\begin{equation}
\label{AP}
V^{'}=-V_{1} \cos(\theta) cos(\frac{\pi z }{a}),
\end{equation}
where  \( V_{1} \) is assumed to be positive.
This potential
prefers  orientation of molecules \( \Theta =0 \) for even layers
z=2 (etc.) and \( \Theta = \pi \) for odd layers z=1 (etc.)
The total effective potential is composed from the basic part
(\ref{EP}) and from the additional anisotropy part (\ref{AP})
\begin{equation}
\label{TP}
V_{T} = - V_{0} ((\Sigma+ \sigma \alpha \cos(\frac{2\pi z}{a})) \frac{1}{2}
(3 \cos^{2}(\Theta)-1)
+ S \gamma \cos(\theta) cos(\frac{\pi z }{a}))
\end{equation}
where
\[ \gamma \equiv \frac{V_{1}}{V_{0}}. \]
The total effective potential depends on the orientation of
molecules \( \theta \) within the layers.

\subsection{Thermodynamics}
The single  particle effective distribution function has the form
\begin{equation}
\label{DF}
f(\theta,z)= \exp{(-\frac{V_{T}(\theta,z)}{k_{B}T})}
\end{equation}
Selfconsistency conditions are
\begin{eqnarray}
\label{SC}
\Sigma & = & <\frac{3 \cos^{2}(\theta)-1}{2}> \nonumber \\
\sigma & = & <\cos{(\frac{2\pi z}{a})} \frac{3 \cos^{2}(\theta)-1}{2}> \nonumber \\
S & = & < \cos{(\frac{\pi z}{a})} \cos{\theta} > \\
\end{eqnarray}
The free energy F=U-TS, where U is the internal energy, S is the
entropy and T is temperature, may be easily calculated. Here
\begin{equation}
\label{U}
U=-\frac{1}{2} N V_{0} (\Sigma^{2}+\alpha \sigma^{2}+S^{2}\gamma),
\end{equation}
and
\begin{equation}
\label{S}
-TS=N V_{0} (\Sigma^{2}+\alpha \sigma^{2}+S^{2}\gamma) - N k_{B} T
\ln{\frac{1}{2a}\int^{2a}_{0} dz \int^{1}_{0} d(cos(\theta))f(z,
\theta)}.
\end{equation}

\subsection{High temperatures}
High temperature free energy expansion may be easily calculated
from (\ref{U}) and (\ref{S}):
\begin{equation}
\label{FH}
F= \frac{NV_{0}}{2}
[(1-\frac{T_{0 \Sigma}}{T}) \Sigma^{2}+\alpha
(1-\frac{T_{0\sigma}}{T})\sigma^{2}+\gamma
(1-\frac{T_{0S}}{T})S^{2}] + higher_order_terms
\end{equation}
where
\[ T_{0} \equiv \frac{V_{0}}{k_{B}}, \]
\[ T_{0 \Sigma} \equiv \frac{T_{0}}{5} \]
\[ T_{0\sigma} \equiv \frac{T_{0}\alpha}{10} < T_{0 \Sigma} \]
with
\[ 0 \leq \gamma <1, \]
and where
\[ T_{0S} \equiv \frac{T_{0}\gamma}{6} \]
and where
\[  \frac{5\gamma}{6} <1 \]

\subsection{Phase transitions}
Within the McMillan theory presented above there are two
possible sequencies of phase transitions.

If
\[ \frac{5\gamma}{6} > \frac{\alpha}{10} \]
then
\[ I \rightarrow N \rightarrow Sm_{A} \rightarrow Sm(AF) \]
Note that the condition above is equivalent to
\[ \gamma > \frac{6}{25} \exp{-(\frac{\pi r_{0}}{a})^{2}}, \]
which holds for molecules with characteristic length \( r_{0} \)
much larger than the interlayer spacing a, and systems in which
additional anisotropy forces are not negligible.
The phase \( Sm_{A} \) is a new bi-layered phase, the phase
Sm(AF) is of the antiferroelectic type descibed above.
The McMillan theory presented here is not constructed to
describe the ferrielectric phases mentioned in the first part of
the paper.

Another possibility is:

If
\[ \frac{5\gamma}{6} < \frac{\alpha}{10} \]
then
\[ I \rightarrow N \rightarrow Sm(AF)  \rightarrow Sm_{A} \]

Structure and phase transitions in smectic A liquid crystals with
polar and sterical asymmetry are discussed in \cite{BIO} and \cite{dG}
There is a large variety of bi-layered structures with smectic A
type ordering. As far as we know a theory which describes as well
traditional ferro-, antiferro- and ferri- electric phases as
corresponding new bi-layered phases is absent. Results of this
paper may, at least partially, cover this gap.
Antiferroelectric chiral smectic liquid crystals are discussed in 
\cite{FTIIT}. The difference between the \( SmC^{*}_{A} \) and
\( SmC_{\alpha}^{*} \) phases is described in this latter paper and
it is explained by existence and non-existence of the dipole
moment pairs.

\end{document}